\documentclass{article}

\usepackage[a4paper,top=3cm,bottom=2cm,left=3cm,right=3cm,marginparwidth=1.75cm]{geometry}

\usepackage[utf8]{inputenc}
\usepackage{amsmath}
\usepackage[numbers]{natbib}
\usepackage{mathtools}
\usepackage{graphicx}
\usepackage{setspace}
\usepackage{amssymb}
\usepackage{hyperref}
\usepackage[ruled,vlined]{algorithm2e}
\title{Real-time super-resolution mapping of locally anisotropic grain orientations for ultrasonic non-destructive evaluation of crystalline material}
\author{J. Singh$^{1,*}$,  K. M. M. Tant$^{1}$, A. Curtis$^{2}$, A. J. Mulholland$^{3}$\\
\\$^1$Department of Mathematics and Statistics, University of Strathclyde
\\$^{2}$School of Geosciences, University of Edinburgh
\\$^{3}$Department of Engineering Mathematics, University of Bristol

\\$^*$Corresponding author: jonathan.singh@strath.ac.uk}

\begin{document}
\maketitle

\begin{abstract}
Estimating the spatially varying microstructures of heterogeneous and locally anisotropic media non-destructively is necessary for the accurate detection of flaws and reliable monitoring of manufacturing processes. Conventional algorithms used for solving this inverse problem come with significant computational cost, particularly in the case of high dimensional non-linear tomographic problems. In this paper, we propose a framework which uses deep neural networks (DNNs) with full aperture, pitch-catch and pulse-echo transducer configurations to reconstruct material maps of crystallographic orientation. We also present the first ever application of generative adversarial networks (GANs) to achieve super resolution of ultrasonic tomographic images, providing a factor-four increase in image resolution and up to a 50\% increase in structural similarity. The importance of including appropriate prior knowledge in the GAN training dataset to increase inversion accuracy is highlighted; known information about the material's structure should be present in the training data. We show that after a computationally expensive training process, the DNNs and GANs can be used in less that one second (0.9 seconds on a standard desktop computer) to provide a high resolution map of the material's grain orientations. 

\end{abstract}


\section{Introduction}

Ultrasonic non-destructive evaluation (NDE) is widely used across a number of industries including aerospace, nuclear, and oil and gas. The technique involves the generation, transmission and reception of high-frequency mechanical waves through a component \cite{blitz1995ultrasonic}. An image of the component's interior is then generated via post processing of this data to aid in the detection of any internal defects \cite{lines1998rapid}. 
Conventional ultrasonic imaging algorithms within NDE typically assume that the material that is being inspected is isotropic and homogeneous. However, metals can develop locally anisotropic and heterogeneous microstructures, particularly when they are subjected to extreme thermal cycles, such as those present in welding and additive manufacturing processes \cite{chassignole2000characterization,rodrigues2019wire,wang2019correlation}. Conventional ultrasonic imaging algorithms which assume homogeneity or isotropy can fail to focus the energy correctly in the image domain in such cases and are therefore unreliable \cite{nageswaran2009microstructural,tant2018transdimensional,zhang2012monte}. Algorithms which incorporate \textit{a priori} information about a material's spatially varying properties significantly improve the accuracy of defect characterisation \cite{tant2018transdimensional}. 

In recent years, much effort has been expended on generating material property maps non-destructively using tomographic inversion, where material properties such as wave speed, or microstructural descriptors such as grain orientation, are estimated from the scattered wave field data recorded at the surface of an object. A wide range of advanced tomographic algorithms are used across geophysics  \cite{aki1976determination,aki1977determination,bodin2009seismic,lebedev2008global,virieux2009overview,zhang2020variational,zhao2020bayesian}, bio-medicine \cite{duric2005development} and NDE \cite{capineri1992time,khairi2019ultrasound,tant2018transdimensional,tant2020effective}. A common approach is to use iterative methods to improve the fit of the measured data to forward modelled data which depend on an estimate of the material map. They sample potential material maps from some multi-dimensional parameter space, solve a forward problem for each new material property map, and update the estimated map to improve the data fit \cite{khairi2019ultrasound}. In the case of probabilistic sampling frameworks (for example, those built around Markov chain Monte Carlo methods \cite{tant2020effective,zhang20183}), there is the added benefit of extracting uncertainty information on the parameter estimates, facilitating valuable uncertainty quantification studies. 
Although these algorithms have demonstrated impressive results in reconstructing wave speed and grain orientation maps, they are computationally demanding, often requiring the storage of large sample sets and compute times of several hours to several weeks. 
This poses a problem for the NDE community, where there is an increasing demand for the monitoring of dynamical processes employed during manufacturing, for example in welding and additive manufacturing processes \cite{javadi2019ultrasonic,javadi2020process}, and so it is desirable to carry out inspection in real-time.   

Machine learning shows strong potential to solve material characterisation inverse problems rapidly \cite{earp2020probabilistic}. Specifically, we focus on the use of deep neural networks (DNNs), which can approximate any non-linear relationship between two parameter spaces, given a sufficiently large set of training data (pairs of dependent and corresponding independent parameters \cite{bishop1995neural}). The training of a DNN is computationally expensive. However, the training process is only performed once prior to using a DNN, and a trained network can be used effectively in real time without the need for high-performance computing.

Inversion methods based on DNNs have become increasingly popular for tomographic imaging of isotropic material properties, particularly in geophysics \cite{araya2018deep,bianco2018travel,cao2020near,earp2020probabilistic,moya2010inversion} and bio-medicine \cite{antholzer2019deep,yoo2019deep}. However, DNNs have not yet been implemented for tomographic reconstruction of anisotropic material properties.
Although various deep learning algorithms have been used to solve inverse problems in NDE, for example, to predict material fatigue behaviour \cite{amiri2020applications}, to augment ultrasonic data \cite{virkkunen2021augmented}, and for ultrasonic crack characterisation \cite{pyle2020deep} and crack detection using image recognition \cite{cha2017deep,huang2018deep}, the use of DNNs for tomography has yet to be explored in this context. 

In addition to DNNs, generative adversarial networks (GANs) have more recently been applied to various computer vision tasks, including achieving super-resolution with upscaling by up to a factor of four \cite{ledig2017photo}, colourisation \cite{guadarrama2017pixcolor}, segmentation and labelling \cite{isola2017image}. This family of algorithms has strong potential to improve image resolution, and has been used increasingly in remote sensing \cite{jiang2019edge} and X-ray tomography \cite{you2019ct}, however there has been no application of GANs in NDE to produce ultrasonic tomographic images. 

In this paper, we present the first DNN framework for rapid, non-linear two-dimensional tomography of heterogeneous and locally anisotropic materials. The datasets used for the tomographic inversion are the arrival times of ultrasonic waves which have been transmitted and received by an array of sensors on the exterior of the component. The examples shown are inspired by the NDE of polycrystalline materials but the methodology should naturally extend to other domains, for example imaging anisotropic fibrous tissue \cite{eltony2020measuring,hoffmeister1994effect} or the Earth's subsurface \cite{zhu2017radial}. We compare the network's performance for a range of transducer configurations, model textures and different types of simulated ultrasonic testing data (i.e. we move beyond inverse crime scenarios). A novel GAN-based method for post-processing ultrasound tomographic images to achieve super-resolution with a four-fold upscaling factor is presented, achieving up to 50\% improvement using structural similarity metrics. We define the term super-resolution in the context of image processing, as reconstructing images below the original lengthscale. This is different to an alternative definition often used in physical acoustics, which is to image below the wavelength in the data.  

\section{Method}

\begin{figure}[t]
    \centering
    \includegraphics[width=\textwidth]{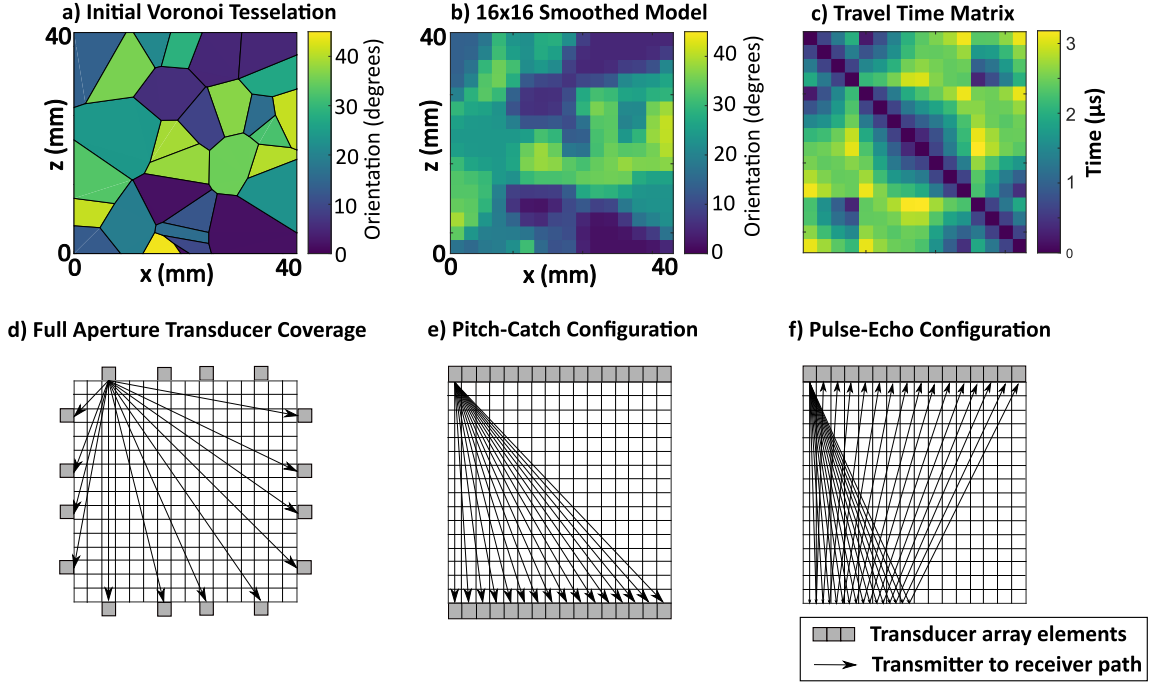}
    \caption{Illustration of the procedure for generating training data: (a) Randomly generated Voronoi tessellation with 30 seeds and random grain orientations ranging between 0$^\circ$ and 45$^\circ$ (other angles are included due to symmetry of the slowness curve). (b) Material map $\mathbf{m}_{16}$ generated by discretising the Voronoi image (a) on a 16x16 grid and then smoothing with a Gaussian kernel. (c) An example travel time matrix $T_{\mathbf{m}_{16}}$ populated by measurements of the time-of-flight between every pair of transducer elements. In this paper, we position transducer elements in three configurations: (d) full aperture coverage by 16 elements (4 on each face), (e) two-sided aperture coverage with 16 transmitting elements at the top of the model with the pitch-catch time-of-flight being measured at 16 receiving elements at the bottom of the model, (f) one-sided aperture coverage, where 16 transducers are positions along the top face and the pulse-echo arrival time of the wave reflecting from the bottom face is measured.        }
    \label{fig:models}
\end{figure}

We employ model-driven deep learning, where a large dataset of simulated material maps and corresponding travel time measurements are used to train a DNN and hence solve the tomographic inverse problem. The forward modelling problem can be denoted as
\begin{equation}
     f(\mathbf{m},\mathbf{s})=\mathbf{T_{m}}, 
\end{equation}
where $f$ is a forward mechanical wave modeling operator, $\mathbf{m}$ is a material model, $\mathbf{s}$ contains the locations of the elements in the ultrasonic transducer array and $\mathbf{T_m}$ is the time-of-flight (ToF) matrix between every pair of array elements. Within each database used for network training, the transducer configuration $\mathbf{s}$ is fixed and therefore $\mathbf{s}$ is omitted in the notation for the ToF matrix $\mathbf{T_m}$. We use deep learning to obtain (or learn) an approximation of $f^{-1}$, which maps the measured data $\mathbf{T_m}$ to a material map $\mathbf{m}$ (i.e., $DNN	\approx f^{-1} $). In this study, the training data consists of two-dimensional material models with spatially varying crystal orientations $\theta(x,y)$ and the travel time matrix $\mathbf{T_m}$ corresponing to each one. To generate models in such a way that the distribution of orientations are randomly assigned but still exhibit some structural correlation, an initial random Voronoi tessellation \cite{senechal1993spatial} with 30 seeds (a set of two-dimensional Cartesian coordinates lying within the domain of interest) is computed and an orientation $\theta$ between $0^\circ$ and $45^\circ$ is randomly assigned to each of the 30 resulting Voronoi regions or cells (Fig.\ref{fig:models}a). We consider only in-plane crystal rotation, and therefore the orientation $\theta$ relates to the orientation of a slowness curve in each cell. This slowness curve plots the reciprocal of velocity in the crystal over a range of incident wave directions \cite{tant2020effective}. The material models used in the training data $\{\mathbf{m}_{16},\mathbf{T_{m_{16}}}\}$ are generated by discretising the Voronoi tessellation into a regularly spaced $16\times 16$ grid and smoothing with a Gaussian kernel (Fig.\ref{fig:models}b, the subscript 16 denotes the model resolution). The smoothing simplifies the inverse problem so that only smooth models are inverted for. To demonstrate that this machine learning approach can be generalised for any locally anisotropic media, the longitudinal group slowness curve is obtained for an arbitrary anisotropic material with a cubic stiffness tensor,  where $c_{11}=256.45$ GPa, $c_{12}=133.5$ GPa and $c_{44}=c_{12}$ and density $\rho = 7874$ kgm$^{-3}$. Three configurations of ultrasonic transducer array locations $\mathbf{s}$ are considered: a full aperture coverage of 16 elements (4 on each face as shown in Fig. \ref{fig:models}d), a two-sided aperture pitch-catch configuration with 16 transmitting elements at the top of the model, with the time-of-flights measured at 16 receiving elements at the bottom of the model (Fig. \ref{fig:models}e), and a one-sided aperture pulse-echo configuration, where 16 elements are positioned along the top face and the travel times of waves reflecting off the bottom face and returning to the transducers on the top face are measured (Fig. \ref{fig:models}f). The measured data is the time-of-flight (ToF) of each propagating wave between each pair of array elements, represented in a ToF matrix $\mathbf{T_{m_{16}}}$ shown in Figure \ref{fig:models}(c).

\subsection{Forward Model Approaches}

Following model parameterisation, an efficient forward model is required for computing the time-of-flight matrix $\mathbf{T_m}$ (Fig.\ref{fig:models}d) corresponding to a grain orientation model $\mathbf{m}_{16}$ for each source-receiver pair. We take two approaches: a semi-analytic model using an anisotropic multi-stencil fast marching method (AMSFMM) algorithm from \cite{tant2020effective}, denoted as $f_{\mbox{FMM}}$, and a finite element analysis (FEA) method, denoted as $f_{\mbox{FEA}}$. 
The AMSFMM incorporates the effects of ray bending due to variations in locally anisotropic grain orientations, and models the travel-time field by solving the Eikonal equation using an upwind finite difference scheme \cite{rawlinson2004wave,sethian1999level,tant2020effective}. This allows the calculation of the shortest travel time between transmitter and receiver locations and the matrix $\mathbf{T_m}^{\mbox{FMM}}$ can be constructed (that is $\mathbf{T_m}^{\mbox{FMM}}=f_{\mbox{FMM}}(\mathbf{m}_{16})$). As wave reflections are not incorporated into the AMSFMM, a different approach is required for the pulse-echo transducer array configuration. In this case, the time of flight between the transmitter and receiver is calculated by the summation of the time of flight between the transmitter to all points along the back-wall and between the receiver and all points along the bottom face. The output of this summation is an array of travel times corresponding to all the reflection points along the bottom face, and the minimum value is taken to be the time of flight for the pulse-echo transducer array configuration.  
The FEA method incorporates more of the underlying physics in the model compared to AMSFMM, as it models full wave propagation including multiple scattering and diffraction. To measure the ToF of the received waves, an amplitude threshold is selected and the time for the recorded wave amplitude to reach this threshold is used as an element of the travel time matrix $\mathbf{T_m}^{\mbox{FEA}}$ (that is $\mathbf{T_m}^{\mbox{FEA}}=f_{\mbox{FEA}}(\mathbf{m}_{16})$). The FEA method is significantly more computationally expensive than the AMSFMM. As a large number of data-model pairs are required to train a deep neural network, the more efficient AMSFMM method is used to generate travel time matrices $\mathbf{T_m}^{\mbox{FMM}}$ for the training data. The more physically realistic FEA generated data is then used to generate data to test the trained networks' performance (see FEA set-up in the Appendix).
A total of 7500 models are generated and the corresponding travel time matrices $\mathbf{T_m}^{\mbox{FMM}}$ are computed using AMSFMM for the training data set.

\subsection{Deep neural network for orientation mapping}
\label{section:dnn}

Deep neural networks (DNNs) are mathematical mappings that emulate the relationship between two parameter spaces \citep{earp2020probabilistic}. Here, we seek a map between the grain orientation models $\mathbf{m}_{16}$ and the corresponding time of flight data $\mathbf{T_m}$ (that is $DNN(T_{16}^{\mbox{FMM}}) = \mathbf{m}_{16}^{\mbox{pred}}$, where the $\mbox{pred}$ superscript denotes the DNN prediction). For each of the transducer configurations $\mathbf{s}$, a different number of travel times are used as input to the neural network. For a full aperture configuration (Fig. \ref{fig:models}d), we have $n$ source-receivers per side of our rectangular domain, and so there are $6n^2$ unique travel times (accounting for source-receiver reciprocity and excluding those between elements which lie on the same side). When $n=4$, a set of 96 travel times is taken from each ToF matrix $\mathbf{T_m}$. For the pitch-catch configuration configuration (Fig. \ref{fig:models}e), all source-receiver paths are unique, therefore the full ToF matrix is used and with $n=16$ there are 256 inputs to the neural network. Finally, for the pulse-echo configuration (Fig. \ref{fig:models}f), when $n$ is even, there are $n^2/2 + n/2$ unique travel times (accounting for source-receiver reciprocity), so when $n=16$, a total of 136 travel times are selected for the ToF matrix.
For network training, both the input travel times and the output orientations are scaled to have zero mean and unit variance. 

\begin{figure}
    \centering
    \includegraphics[width=\textwidth]{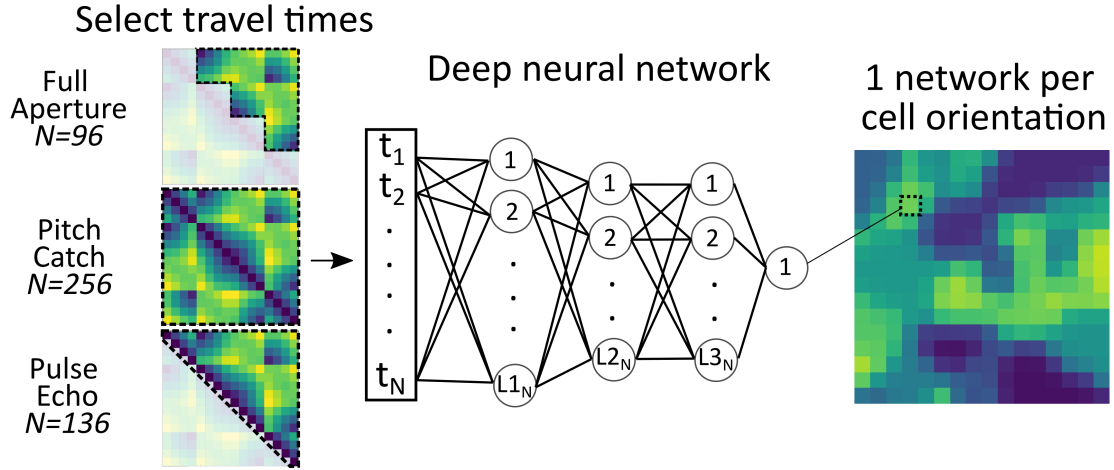}
    \caption{Schematic of the tomography algorithm using deep neural networks (DNNs). First, $N$ travel times are selected from the travel time matrix (where $N=96$, $N=256$ and $N=136$ for the full aperture, pitch-catch and pulse-echo configurations respectively). Travel times are then used as input into the DNN. The DNN consists of 3 hidden layers ($L1$, $L2$ and $L3$) and a final output layer. The nodes are illustrated as circles and the number of nodes in each layer is denoted in the bottom circles ($L1_N$, $L2_N$ and $L3_N$).  Each output corresponds to the crystal orientation of a single pixel in the material map $\mathbf{m_{16}}$ so 256 separate networks are trained in order to predict all pixel orientations. }
    \label{fig:dnn_cartoon}
\end{figure}

We configure three DNNs (corresponding to three transducer configurations), each with five fully connected layers (illustrated in Fig. \ref{fig:dnn_cartoon}), using sigmoid activation functions. The final output layer contains a single node corresponding to the orientation of a single pixel in the imaging domain. Therefore, following the approach of \cite{earp2020probabilistic}, a separate network is trained for each pixel, so for a $16\times16$ resolution image a total of 256 networks are trained. 
The networks are trained using the Adam optimisation algorithm \cite{kingma2014adam}. A description of network hyper-parameters is provided in Appendix \ref{appendix:params}. These hyper-parameters are selected using a stochastic optimisation library \cite{bergstra2015hyperopt} for each network architecture corresponding to different transducer configurations. We use a mean-squared-error (MSE) loss function, given by:
\begin{equation}
    MSE = \frac{\sum^N_{i=1}(\mathbf{m}_{16}^{\mbox{true}}(i)- \mathbf{m}_{16}^{\mbox{pred}}(i))^2}{N},
\end{equation}
where $\mathbf{m}_{16}^{\mbox{true}}$ and $\mathbf{m}_{16}^{\mbox{pred}}$ are the true and predicted grain orientation models, $i$ denotes the pixel index and $N$ is the total number of pixels (for models $\mathbf{m}_{16}$, $N=256$).
A validation data set is created using 20\% of the training data. To avoid over-fitting the network to the training data, the cost function is periodically evaluated over the validation data set, and we implement an early stopping algorithm so that training stops once the validation loss stops decreasing (with a patience of 10 iterations). The time to train 256 separate networks sequentially using \textit{Google Colab} \cite{bisong2019google} free GPUs is approximately 40 minutes, although training could be parallelised to reduce this time if required. Once trained, the compute time of $\mathbf{m}_{16}^{\mbox{pred}}$ is approximately 0.15 seconds per model inversion. 

\subsection{Generative adversarial networks for super resolution}
\begin{figure}
    \centering
    \includegraphics[width=0.9\textwidth]{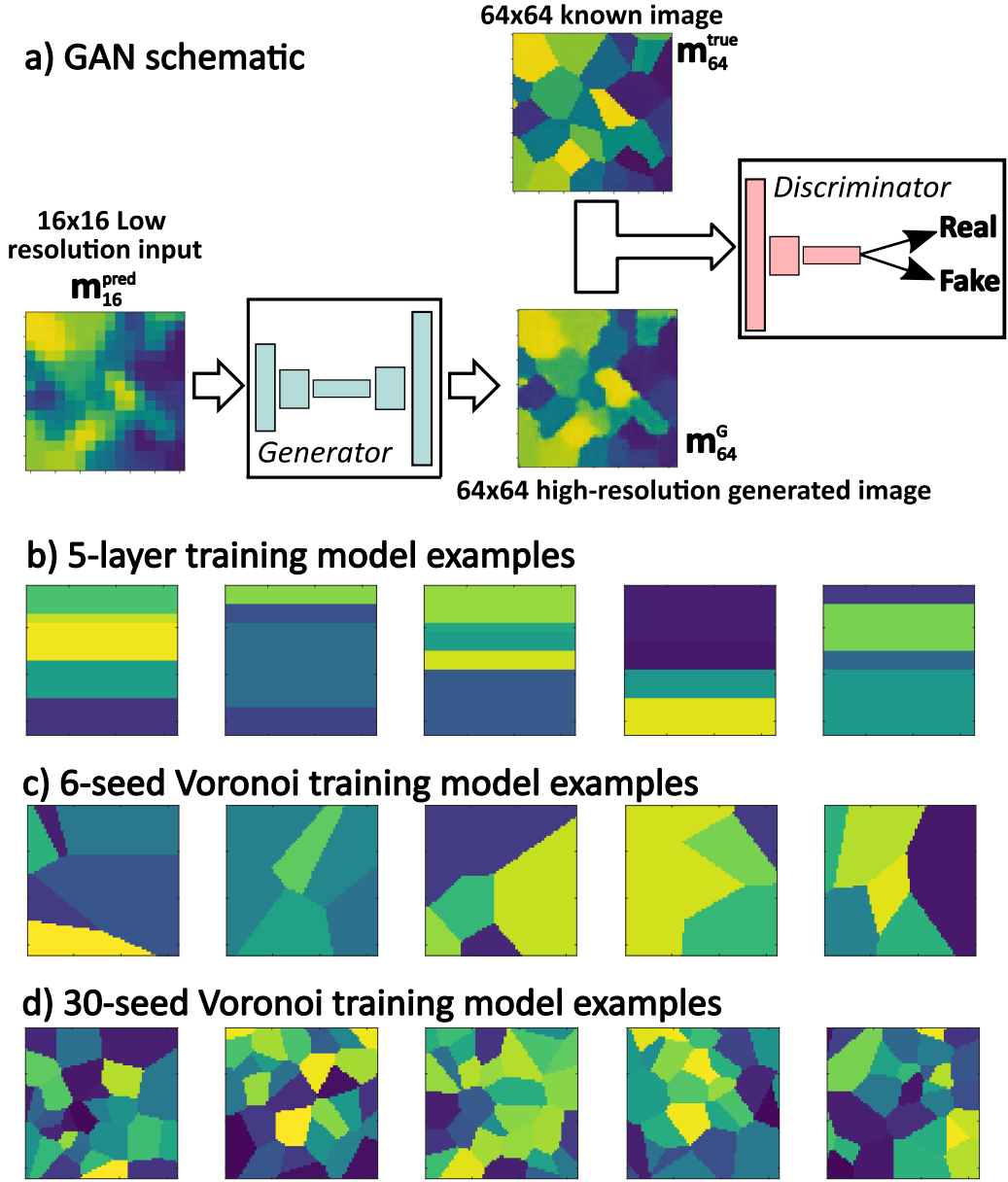}
    \caption{(a) Schematic of the Generative Adversarial Network (GAN) post-processing algorithm for achieving super-resolution material maps. The 16x16 output of the DNN tomography algorithm $\mathbf{m}_{16}$ is input to the generator network, which outputs a 64x64 image $\mathbf{m}_{64}^{G}$. The known 64x64 image $\mathbf{m}_{64}^{T}$, which was used to generate the DNN input data, as well as the output of the generator $\mathbf{m}_{64}^{G}$ are input into the discriminator, which outputs a prediction of which image is generated and which belongs to the training dataset. Three different model data sets are used for GAN training: (b) a layer model with up to 5 layers where positions of the interfaces are random, (c) a random Voronoi tessellation with 6 seeds, and (c) a random Voronoi tessellation with 30 seeds.  }
    \label{fig:gan_cartoon}
\end{figure}
Conditional GANs learn a mapping between two images \cite{isola2017image} and so can be used for post-processing of the DNN tomography output ($\mathbf{m}_{16}^{\mbox{pred}}$) to increase resolution and accuracy. The GAN architecture, as illustrated in Figure \ref{fig:gan_cartoon}(a), consists of two separate trainable networks: a generator ($\mbox{GAN}_G$) and a discriminator ($\mbox{GAN}_D$).

Training a GAN for post-processing the output of the DNN tomography method ($\mathbf{m}_{16}^{\mbox{pred}}$) to achieve an increase in image resolution (super-resolution) requires an additional training data set, where travel time data is generated using higher resolution models (64x64).
The GAN framework assumes some prior knowledge of the structure of the material which is incorporated into the GAN training data. For example, in layered structures such as Carbon fiber reinforced polymers (CFRPs), the training data should include models with locally anisotropic layers; or alternatively models exhibiting crystalline grain structures should be used to train GAN's for cases such as welds, and knowledge on the average grain size could feed into the complexity of the models included in the training data. 
We use three separate training data sets of increasing complexity. The first high resolution model  $\mathbf{m}_{64}^{\mbox{true}}$ consists of up to 5 horizontal layers where the orientation and thickness of each layer is randomly assigned (Fig. \ref{fig:gan_cartoon}b). The second and third are generated by discretising random Voronoi tessellations with 6 and 30 seed locations into a 64x64 grid, as shown in Figures \ref{fig:gan_cartoon}(c) and (d), respectively.
The travel time matrices $\mathbf{T_{m}}_{64}^{FMM}$ are calculated using the AMSFMM algorithm for 2000 models for each of the three data sets, which are input into the DNN tomography algorithm described in the previous section, which outputs a 16x16 predicted model $\mathbf{m}_{16}^{\mbox{pred}}$.
The generator is configured to take the low resolution $\mathbf{m}_{16}^{\mbox{pred}}$ image as input and to output a high resolution 64x64 image $\mathbf{m}_{64}^{G}$ (i.e., 
${\mbox{GAN}}_G (\mathbf{m}_{16}^{\mbox{pred}})=\mathbf{m}_{64}^G  $). 
Here, the generator is a modified U-net \cite{ronneberger2015u} based on fully convolutional layers (see the Appendix for network architecture). The discriminator takes the output of the generator $\mathbf{m}_{64}^{G}$, as well as the known 64x64 high resolution image ($\mathbf{m}_{64}^{\mbox{true}}$) that was used to generate the ToF data, and predicts which image is generated (fake) and which is part of the training data (real). The accuracy of the discriminator prediction can then be established. These competing networks are then trained against each other; in each iteration of training, the accuracy of the discriminator is fed into the loss function of the generator network. The generator seeks to create images $\mathbf{m}_{64}^{G}$ that decrease the discriminator accuracy meaning that $\mathbf{m}_{64}^{G}$ cannot be discriminated from the reference training data $\mathbf{m}_{64}^{\mbox{true}}$. Following the training process, the generator can be used to map from 16x16 images to 64x64 resolution images.

\section{Results}
\subsection{DNN Results}
\begin{figure}
    \centering
    \includegraphics[width=\textwidth]{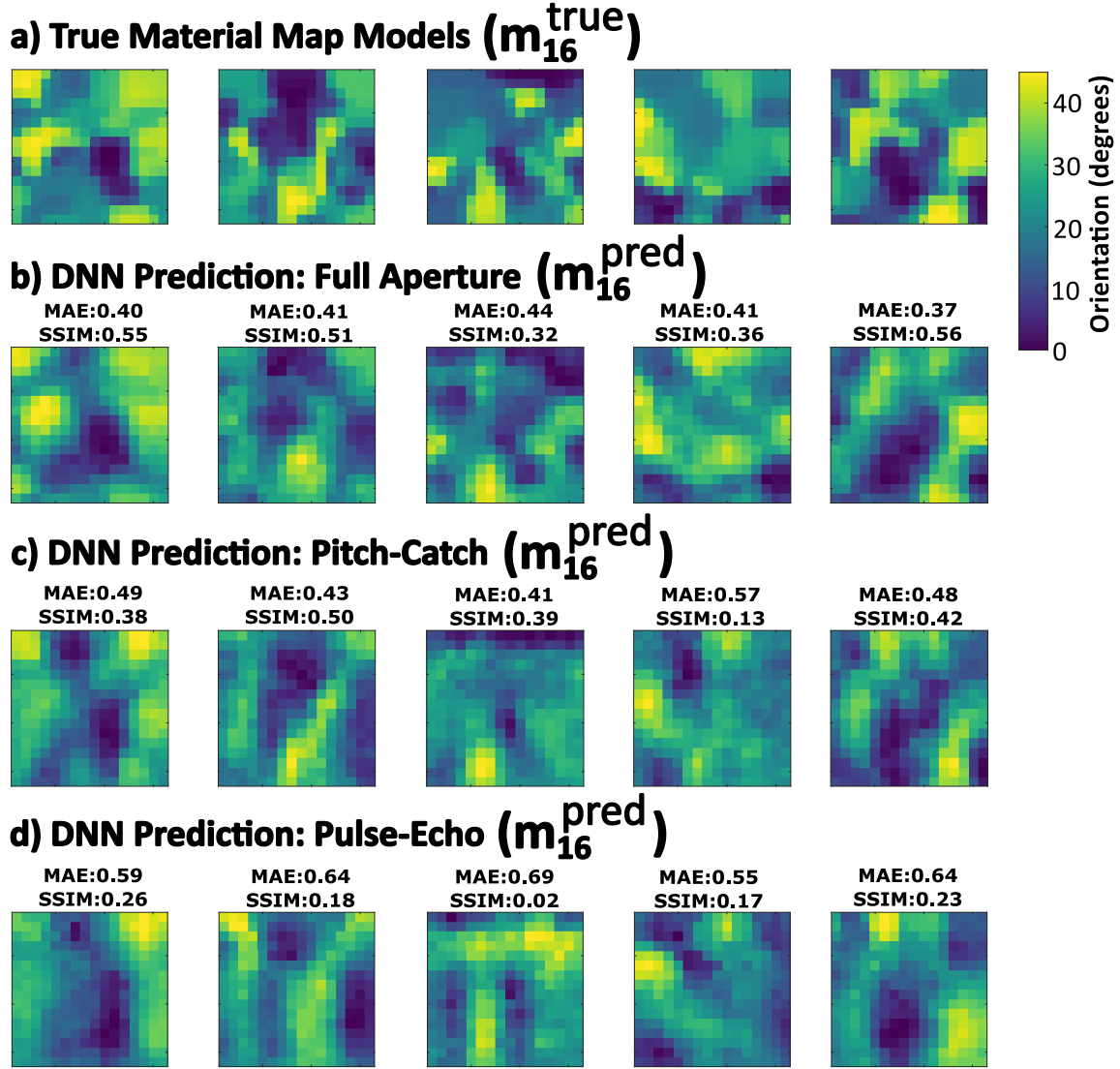}
    \caption{(a) True material orientation maps $\mathbf{m}^{\mbox{true}}_{16}$ from the test data set and corresponding predicted models $\mathbf{m}^{\mbox{pred}}_{16}$ using the DNN tomography algorithm for (b) full aperture, (c) pitch-catch, and (d) pulse-echo transducer array configurations. The associated mean absolute errors (MAE) and the structural similarity index metrics (SSIM) between the predicted and true maps are labelled above each reconstructed material map.}
    \label{fig:dnn_results}
\end{figure}
Following the training of the fully connected DNN, we predict material maps  $\mathbf{m}_{16}^{\mbox{pred}}$ using the three transducer array configurations shown in Figure \ref{fig:models} following
\begin{equation}
     \mathbf{m}_{16}^{\mbox{pred}}= DNN(\mathbf{T_m}^{\mbox{FMM}}),
\end{equation}
where $\mathbf{T_m}^{\mbox{FMM}}$ is test data which has not been used in the network training process. The test data are generated following the same protocol as for the training data, using smoothed Voronoi models $\mathbf{m}_{16}$ and the AMSFMM algorithm to generate a total of 200 test models and data. Comparisons of the true models $\mathbf{m}^{\mbox{true}}_{16}$ with the predicted models $\mathbf{m}^{\mbox{pred}}_{16}$ using the DNN and with full aperture, pitch-catch and pulse-echo transducer array configurations are shown in Figure \ref{fig:dnn_results}. We use two metrics for comparing predicted models with the true models: the mean absolute error (MAE), which is a scalar value (MAE$\geq$ 0, where MAE=0 describes a perfect prediction),  and the structural similarity index measure (SSIM) \cite{wang2004image} (-1$\leq$SSIM$\leq$1, where SSIM=1 describes a perfect prediction). The SSIM incorporates the similarity of three independent parameters: image luminescence, contrast and structure (see the Appendix). These values are calculated with orientations that are scaled to have zero mean and unit variance. Note that \textit{lower} values of MAE indicate higher similarity between the true and predicted models, whereas \textit{higher} values of SSIM indicate higher image similarity.

\begin{figure}
    \centering
    \includegraphics[width=\textwidth]{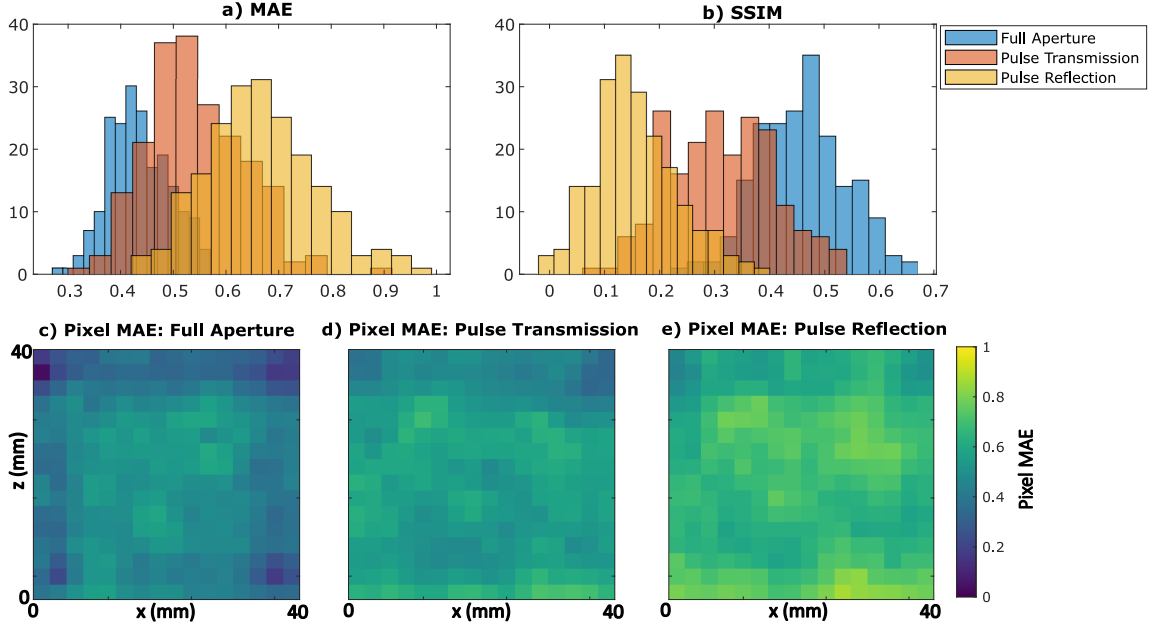}
    \caption{Histogram of (a) mean absolute error (MAE) and (b) structural similarity index metric (SSIM) for DNN prediction on 200 test models using different transducer array configurations. Lower values of MAE and higher values of SSIM suggest a good tomographic reconstruction. The lower panels show the mean absolute error (MAE) for each pixel averaged across the 200 test models for (c) full aperture, (d) pitch-catch and (e) pulse-echo transducer array configurations.}
    \label{fig:dnn_results_errors}
\end{figure}

In all cases, the predicted material property maps resemble the true orientation maps, predicting the magnitude and location of areas with similar orientations. The DNN predictions with a full aperture experimental configuration (Fig. \ref{fig:dnn_results}b) perform the best (lower MAEs and higher SSIMs) and predictions made using the pulse-echo configuration perform the worst (higher MAEs and lower SSIMs). The histograms of MAE and SSIM values for the 200 test models are shown in Figures \ref{fig:dnn_results_errors}(a) and (b). The distributions of the pixel mean absolute error (averaged for each pixel across the 200 models) are shown for each transducer array configuration in Figures \ref{fig:dnn_results_errors}(c-e), showing that reconstruction accuracy generally decreases (increasing pixel MAE) in the central region of the domain and with distance from the transmitting element transducer array.

\begin{figure}[t]
    \centering
    \includegraphics[width=0.9\textwidth]{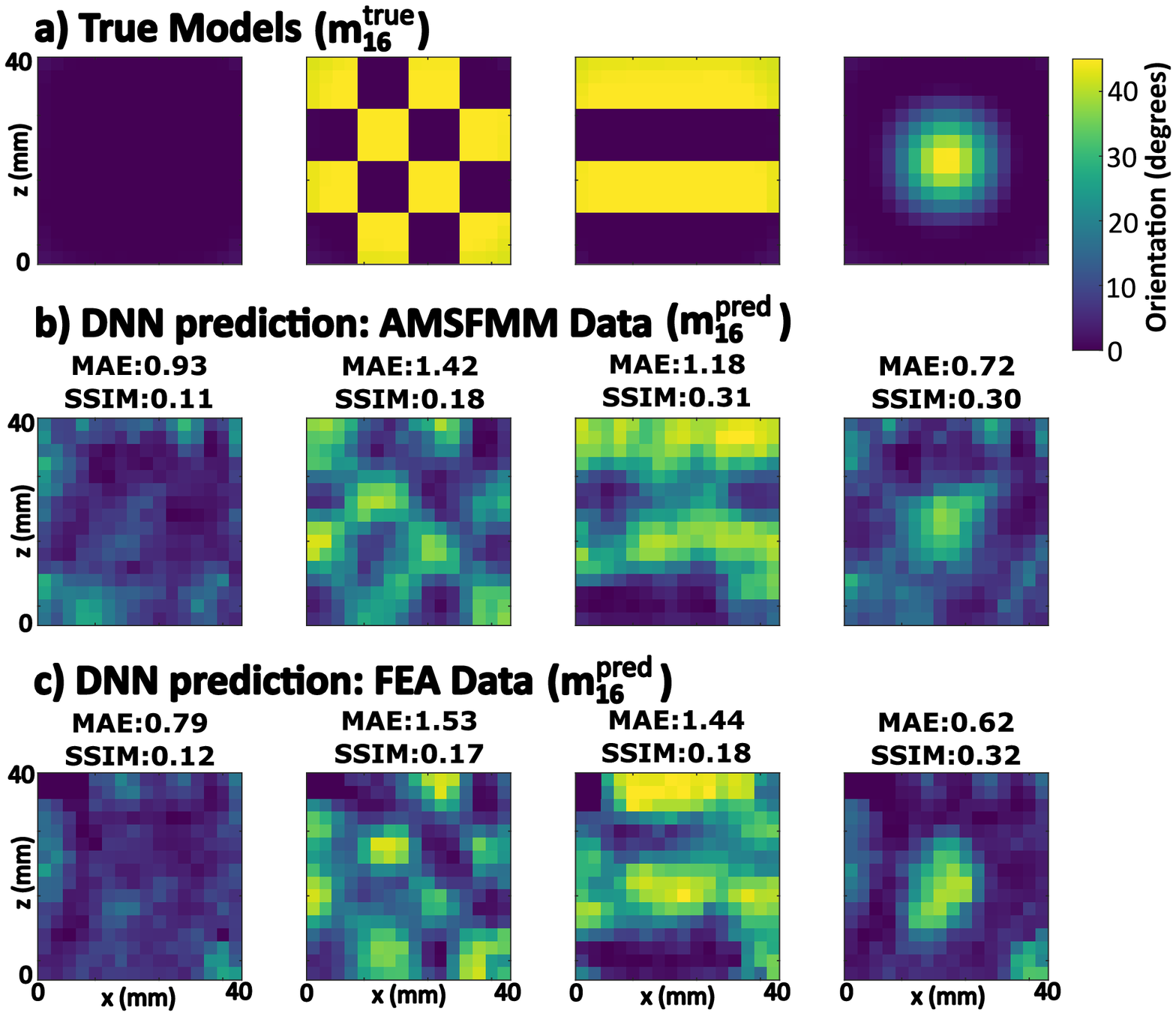}
    \caption{(a) True material maps ($\mathbf{m}_{16}^{\mbox{true}}$). The predicted models using the AMSFMM trained DNN tomography algorithm with (b) $\mathbf{m_{16}}^{\mbox{pred}}=DNN(\mathbf{T_m}^{\mbox{FMM}}$) and (c) $\mathbf{m_{16}}^{\mbox{pred}}=DNN(\mathbf{T_m}^{\mbox{FEA}}$).}
    \label{fig:fea_results}
\end{figure}

So far, the same mathematical model has been used for both the training data and the test data (a so called inverse crime \cite{wirgin2004inverse}), and this is not a sufficient challenge of the methodology \cite{kelly2017deep}.
We therefore now use a different mathematical model to test the trained DNN. One further additional challenge is to generate material maps using a different method from that used in the training data; so not originating from Voronoi diagrams. The material maps in Figure \ref{fig:fea_results}(a) show a range of structures including a homogeneous model, a checkerboard structure, a layered structure and a single circular anomaly; all of which are significantly dissimilar from the textures and structures found within the training data. 
The FEA method is used to generate ToF data $\mathbf{T_m}^{FEA}$ using a full aperture transducer array configuration, which is then input into the DNN to predict the grain orientation map $\mathbf{m}_{16}^{\mbox{pred}}$
\begin{equation}
     \mathbf{m}_{16}^{\mbox{pred}}= DNN(\mathbf{T_m}^{\mbox{FEA}}).
\end{equation}

The predicted material maps $\mathbf{m}_{16}^{\mbox{pred}}$ shown in Figures \ref{fig:fea_results}(b) and (c) show similar results using $\mathbf{T_m}^{\mbox{FEA}}$ and $\mathbf{T_m}^{\mbox{FEA}}$ time of flight data. In the cases of the homogeneous model and the single circular anomaly the results using $\mathbf{T_m}^{\mbox{FEA}}$ are slightly improved (lower MAE). The similarity of results between the two data types indicates that the DNN is robust to changes in different data simulation methods and to the noise in the FEA dataset associated with the identification of travel times. The presence of this additional noise does not appear to have a significant effect on the changes in measured travel time due to anisotropy, therefore the inversion remains accurate.    
The accuracy of the predicted models is lower where the material maps exhibit different textures to those used in the training data; compare the MAE and SSIM values in Figure \ref{fig:fea_results}(c) with those in Figure \ref{fig:dnn_results}(b). The higher accuracy of the results in Figure \ref{fig:dnn_results}(b) highlights that the texture of the target application material for the DNN tomography algorithm should be included as far as possible in the training dataset.

\subsection{GAN Results}

\begin{figure}[h]
    \centering

    \includegraphics[width=0.74\textwidth]{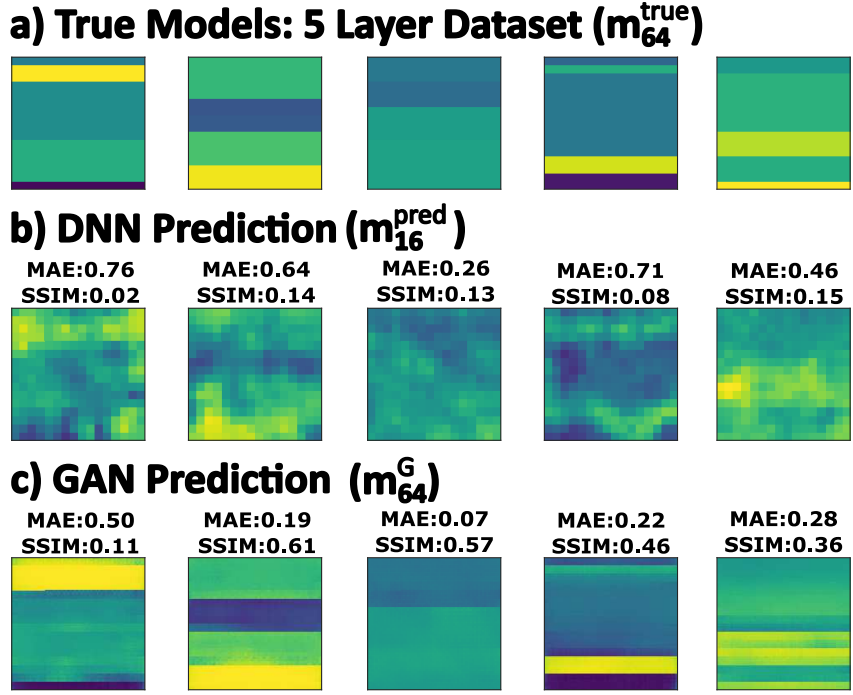}
    \caption{(a) True high resolution (64x64) grain orientation maps $\mathbf{m}_{64}^{\mbox{true}}$ consisting of 5 horizontal layers, (b) 16x16 resolution DNN tomography output ($\mathbf{m}_{16}^{\mbox{pred}}$), and (c) 64x64 GAN output 
    $\mathbf{m}_{64}^{G}$. For row (b), the MAE and SSIM are calculated on an upscaled image to $64\times64$ resolution using nearest neighbour interpolation. Note the significant improvements in MAE and SSIM using the GAN methodology and the clear improvements in reconstructing a layered structure. }
    \label{fig:gan_grid_5lay}
\end{figure}

\begin{figure}
    \centering
    \includegraphics[width=0.74\textwidth]{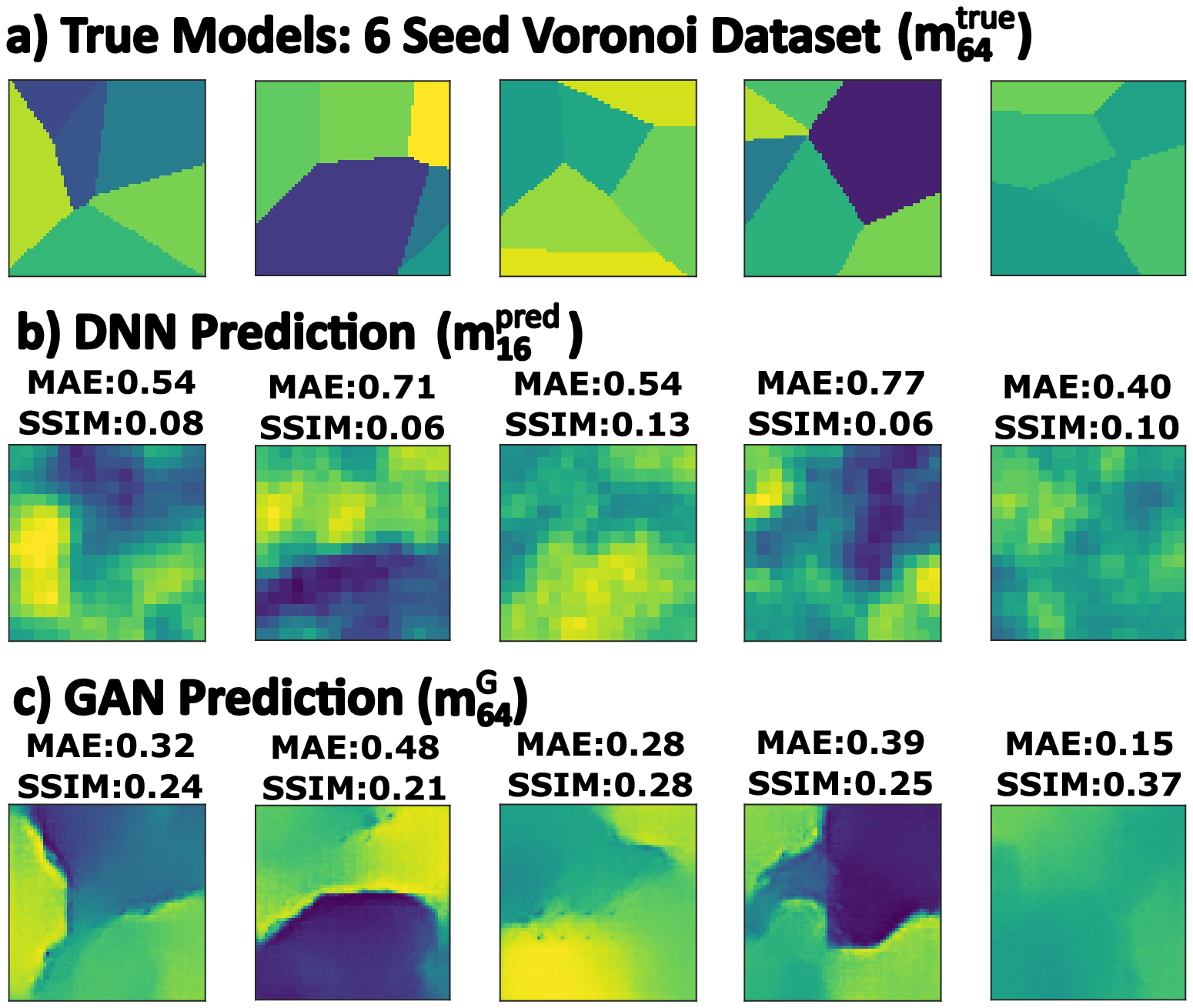}
    \caption{(a) True high resolution (64x64) grain orientation maps $\mathbf{m}_{64}^{\mbox{true}}$ consisting of a Voronoi Tesselation with 6 seeds, (b) 16x16 resolution DNN tomography output ($\mathbf{m}_{16}^{\mbox{pred}}$), and (c) 64x64 GAN output 
    $\mathbf{m}_{64}^{G}$. For row (b), the MAE and SSIM are calculated on an upscaled image to $64\times64$ resolution using nearest neighbour interpolation. Note the significant improvements in MAE and SSIM using the GAN methodology and the clear improvements in reconstructing piecewise constant structures.}
    \label{fig:gan_grid_6seed}
\end{figure}

\begin{figure}
    \centering
    \includegraphics[width=0.74\textwidth]{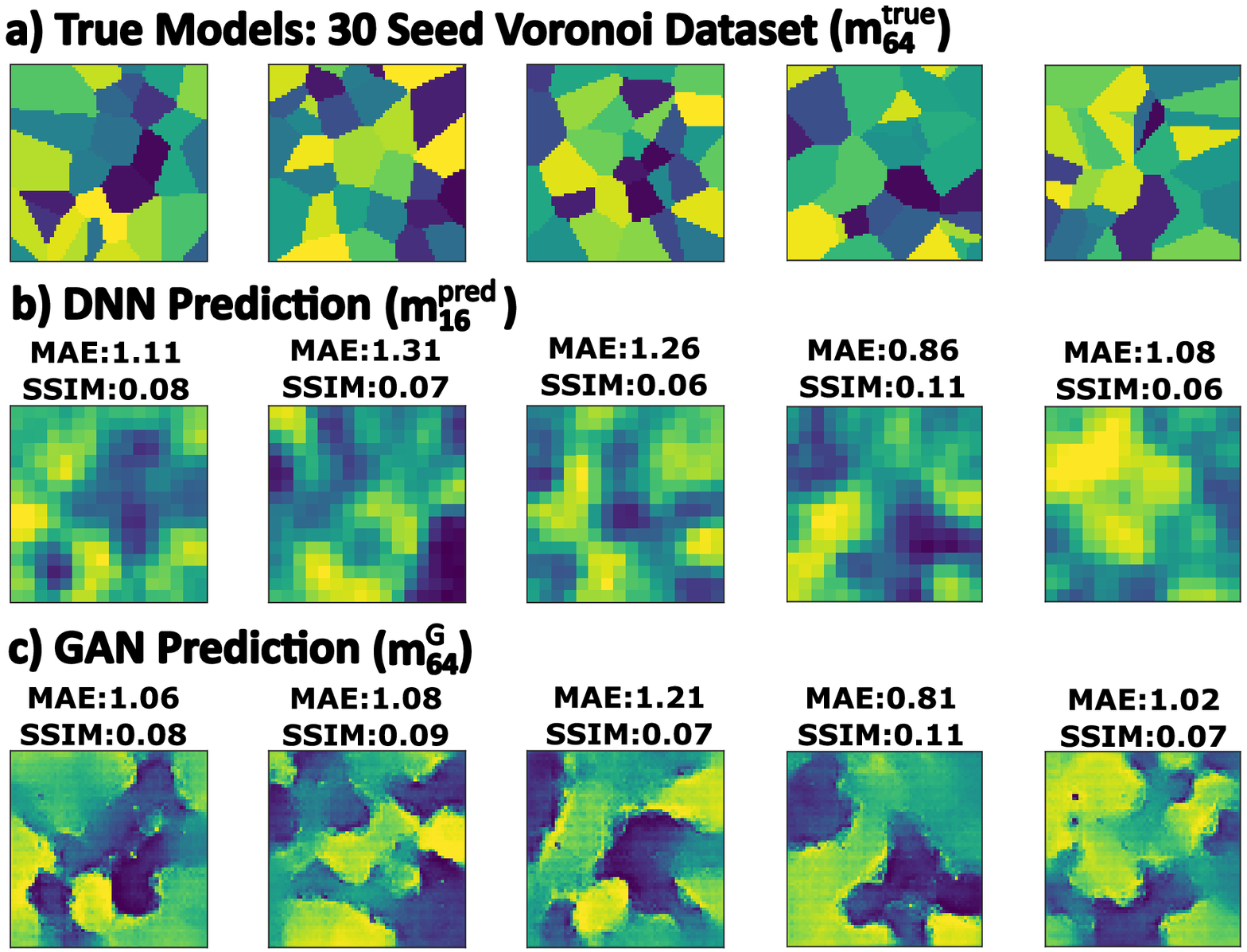}
    \caption{(a) True high resolution (64x64) grain orientation maps $\mathbf{m}_{64}^{\mbox{true}}$ consisting of a Voronoi Tesselation with 30 seeds, (b) 16x16 resolution DNN tomography output ($\mathbf{m}_{16}^{\mbox{pred}}$), and (c) 64x64 GAN output 
    $\mathbf{m}_{64}^{G}$. For row (b), the MAE and SSIM are calculated on an upscaled image to $64\times64$ resolution using nearest neighbour interpolation. Note the marginal improvements in MAE and SSIM using the GAN methodology.}
    \label{fig:gan_grid_30seed}
\end{figure}

\begin{figure}[t]
    \centering
    \includegraphics[width=0.65\textwidth]{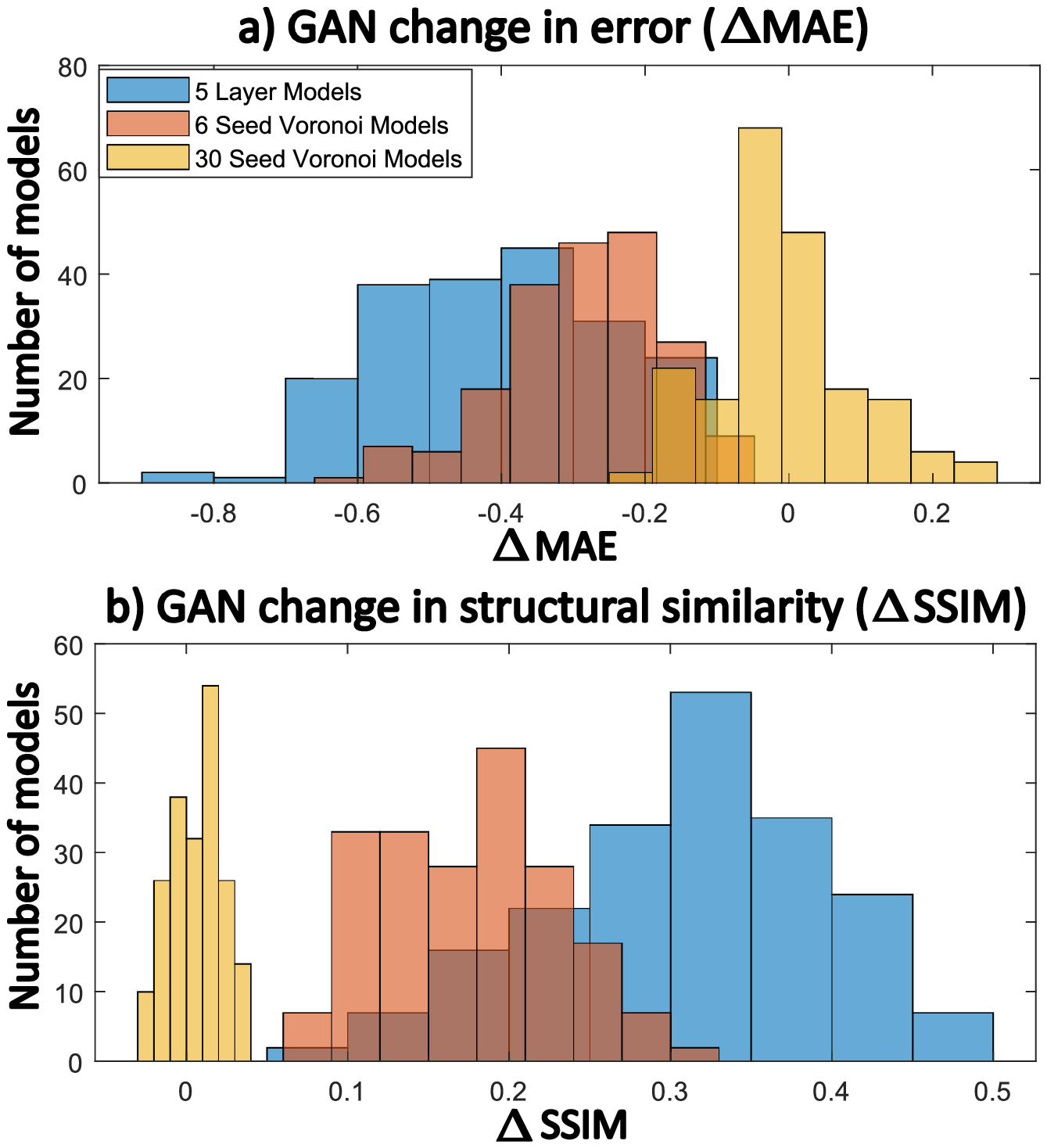}
    \caption{Histograms showing the change in (a) the mean absolute error (MAE) and (b) the structural similarity index measure (SSIM) when using a GAN to post-process 200 low resolution DNN tomography outputs ($\Delta MAE = MAE_{GAN}-MAE_{DNN}$ and $\Delta SSIM = SSIM_{GAN}-SSIM_{DNN}$).}
    \label{fig:gan_hists}
\end{figure}

Three GANs are trained using the layered, 6-seed Voronoi and 30-seed Voronoi models $\mathbf{m}_{64}^{\mbox{true}}$, and 200 additional models per GAN are used for testing, of which 5 are shown in Figures \ref{fig:gan_grid_5lay}(a), \ref{fig:gan_grid_6seed}(a) and \ref{fig:gan_grid_30seed}(a), respectively. The AMSFMM method is used to compute travel time data ($\mathbf{T_m}^{\mbox{FMM}}$) using a full aperture transducer array configuration, which are input into the trained DNN (as used for the generation of DNN predictions in Fig. \ref{fig:dnn_results}b).
The DNN predicted outputs $\mathbf{m}_{16}^{\mbox{pred}}$ are shown in Figures \ref{fig:gan_grid_5lay}(b), \ref{fig:gan_grid_6seed}(b) and \ref{fig:gan_grid_30seed}(b) and the GAN outputs $\mathbf{m}_{64}^G$ in Figures \ref{fig:gan_grid_5lay}(c), \ref{fig:gan_grid_6seed}(c) and \ref{fig:gan_grid_30seed}(c) for the layered, 6-seed Voronoi and 30-seed Voronoi models, respectively. In order for image comparison with MAE and SSIM, the 16x16 resolution DNN outputs are upscaled to 64x64 resolution using nearest neighbour interpolation. Histograms of the changes in MAE ($\Delta MAE  = MAE_{GAN}-MAE_{DNN}$) and SSIM ($\Delta SSIM=SSIM_{GAN}-SSIM_{DNN}$) when using a GAN to post-process the DNN tomography outputs are shown in Figure \ref{fig:gan_hists}.

For the 5 layer models (Fig. \ref{fig:gan_grid_5lay}), the GAN predictions are significantly more accurate compared to DNN predictions, offering large improvements in MAE (decrease up to $\Delta MAE=-0.85$) and SSIM (increase up to $\Delta SSIM=0.5$). The GAN successfully learns to generate horizontal (layered) structures, so very little horizontal variation exists in the GAN predictions. The reconstructed grain orientation maps from the GAN exhibit discontinuous grain boundaries and piecewise constant orientations for each layer, compared to the smooth spatially varying DNN tomography outputs (Fig. \ref{fig:gan_grid_5lay}b). 
The GAN also performs well for the 6-seed Voronoi tessellation models (Fig. \ref{fig:gan_grid_6seed}), where reconstructed grain orientation maps from the GAN exhibit discontinuous, piecewise constant orientations for each grain. The GAN improves MAE and SSIM in all cases, however there is slight blurring across some grain boundaries. The GAN results for the 30-seed Voronoi tessellation models (Fig. \ref{fig:gan_grid_30seed}) exhibit stronger blurring across grain boundaries. While the GAN prediction are texturally more similar to the true models (piecewise constant and discontinuous regions), the distributions of $\Delta MAE$ and $\Delta SSIM$ in Figure \ref{fig:gan_hists} show the GAN offers only marginal improvements in reconstruction accuracy, and in some cases the accuracy decreases when using the GAN ($\Delta MAE>0$ and $\Delta SSIM<0$). The difference between the 6-seed and 30-seed Voronoi models is in the model complexity due to smaller individual grains in the 30-seed models. In these models, multiple grains can fit into a single pixel of a low resolution DNN tomography image, resulting in a loss of spatial information that the GAN cannot fully recover.
These results show that a GAN can be used for post-processing tomography results to improve reconstruction accuracy and image resolution, particularly when prior information regarding the spatial distribution of the material map is known (e.g., if the sample is known to be layered, or similarly well-structured) and the spatial distribution is simple.   

\section{Discussion}
\label{section:discussion}

The framework presented includes several stages: (1) the generation of training data using the AMSFMM method, (2) training of the DNN, and (3) training of the GAN. However, each of these stages only need be performed once. Thereafter, the DNN and GAN can be used in effectively real time ($<1$ second). Here, the time for generating 7500 ToF matrices $T_\mathbf{m}^{\mbox{FMM}}$ was approximately 1 hour, for training the DNN was approximately 40 minutes (until convergence), and for training the GAN was approximately 8 hours (using \textit{Google Colab} GPUs \cite{bisong2019google}). 
It is clear that when repeated material map reconstructions are desired, as is the case for NDE monitoring purposes, the deep learning framework excels in its ability to provide real time results. 

The benefits of real time inversions comes at the expense of a few limitations that are yet to be overcome in the current work.  Firstly, the DNN is trained with a constant transducer configuration, so a trained DNN cannot be generally extended to changes in relative transducer locations. This is not a problem for many applications in NDE, as the transducer arrays are rigid and fixed, and the test sample geometries do not change through time. However, limited network flexibility may be problematic in cases where the configuration changes, such as in-process monitoring of additive manufacturing: during the building process, the shape of the sample changes therefore the distribution of transducer elements also changes. One solution is to train many DNNs for all the possible transducer configurations throughout the building process, however this would require a significantly expensive training process. Another solution, proposed in \cite{earp2020probabilistic}, is to train more flexible networks that account for missing data by augmenting the training data set with additional input samples taken from additional transducer locations. Travel times in the ToF matrix can be set to zero to indicate a transducer is not used for a particular transducer configuration, and then the trained network can invert using multiple configurations. 

\begin{figure}[t]
    \centering
    \includegraphics[width=0.74\textwidth]{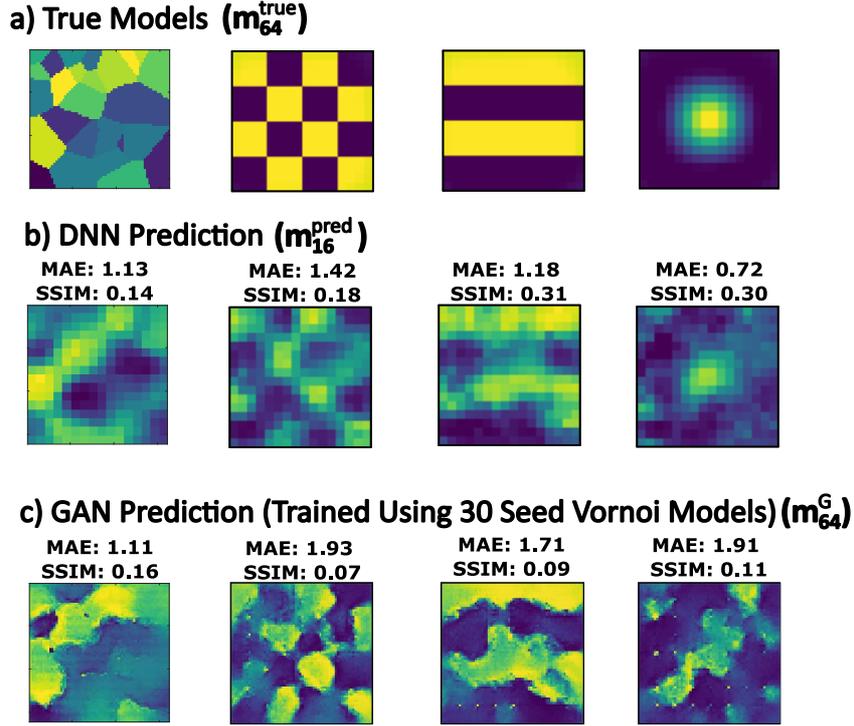}
    \caption{(a) True high resolution (64x64) grain orientation maps $\mathbf{m}_{64}^{\mbox{true}}$, (b) 16x16 resolution DNN tomography output using AMSFMM generated data ($\mathbf{m}_{16}^{\mbox{pred}}$), and (c) 64x64 GAN output 
    $\mathbf{m}_{64}^{G}$. For row (b), the MAE and SSIM are calculated on an upscaled image to $64\times64$ resolution using nearest neighbour interpolation. It can be seen that the similar MAE and SSIM values result for the Voronoi diagram arise since this type of texture was used in the training data of the DNN and GAN. However, the GAN performs significantly worse in the cases where the material texture is not part of the training data in columns 2,3 and 4.}
    \label{fig:gan_grid_shapes}
\end{figure}

The GAN is also limited in its applicability. This is highlighted when a trained GAN is used to invert for textures that are dissimilar to those found in the training data. This can be seen in Figure \ref{fig:gan_grid_shapes}, where the GAN trained on the 30-seed Voronoi models is applied to the DNN prediction of the checkerboard, layered and circular inclusion material models as well as a 30-seed Voronoi model for reference. There are significant decreases in SSIM and increases in MAE when using the GAN on the models with dissimilar textures to the Voronoi models. This highlights the importance of the training data used in network training and suggests the GAN should only be used if prior knowledge of the material is known and the expected textures are present in the training data. Training a GAN with a much broader training data set, for example including all of the layered, 6-seed and 30-seed Voronoi models in the same training data set would allow for more general application of the GAN where less prior knowledge of the material is known. We leave this for future work. 
 
Where real time inversions are not required, more computationally expensive tomography algorithms can be implemented. Algorithms such as the rj-MCMC \cite{tant2018transdimensional} offer more information including an estimate of the uncertainty of the tomography results. A place for rapid deep learning-based tomography still exists within this framework as it can provide a fast, coarse initial model which can be used a a starting point for more sophisticated algorithms. Additionally a GAN can be used in post-processing any tomographic image. Often linearised image methods are often regularised and hence predict smoother structures than are expected to exist in the true medium, therefore a GAN can be trained to upscale resolution and sharpen these images. 
Even where the GAN provides marginal improvements to the DNN tomography results, the GAN output models exhibit discontinuous boundaries. It can be important that such boundaries are present in tomography algorithms where entire waveforms are modelled and matched to the recorded waveforms  (that is, full waveform inversion \cite{virieux2009overview}).
A GAN might also be extended to take the full waveform as an input, though this would require expensive FEA modelling to generate the training data, so that all internal reflections are modelled.

\section{Conclusion}
We present a deep learning based framework for the real time tomographic reconstruction of spatially varying crystal orientations in locally anisotropic media using ultrasonic array time-of-flight data. We train a series of deep neural networks (DNNs) using 7500 models in a training data set, to accurately reconstruct orientation maps using full aperture, pitch-catch and pulse-echo transducer array configurations. We present the first application of generative adversarial networks (GANs) on ultrasonic tomographic data, where a series of GANs are trained with three sets of training data exhibiting increasing levels of complexity in the model textures. The GAN takes the low resolution DNN output and upscales the resolution by a factor of four.  
We show that prior information used to create the training data for both the DNN and the GAN are important factors in providing accurate estimations of the orientation maps. Using the methods presented unlocks a wide range of potential applications for ultrasonic monitoring, allowing for faster and more accurate detection of flaws and in-process inspection during manufacturing. 

\section{Appendix}
\subsection{Finite element analysis}

We implement a finite element simulation of elastic wave propagation in anisotropic media using OnScale \cite{onscale}. We apply absorbing boundary conditions on all sides of the domain so energy continues past boundaries with no reflections. We use Ricker wavelets with central frequencies of 1 MHz as the source-time function, and apply pressure loads following the full aperture transducer array configuration as shown in Figure \ref{fig:models}(d). The values for the finite element node spacing ($\Delta x, \Delta y$) are selected to ensure spatial stability conditions following $\Delta x, \Delta y = \frac{\lambda}{15}$, where $\lambda$ is the shortest wavelength in the domain. 

Following the simulation for each transmitting array element, the travel time to each receiving transducer is automatically picked by selecting the time for arriving energy to increase above a threshold. This threshold is taken to be 2\% of the peak displacement in the recorded signal.

\subsection{Network Architectures}
\label{appendix:params}

The deep neural networks (DNNs) are trained using 5 layers, where each node receives an input from every node in the previous layer and a sigmoidal activation function. The number of nodes in each layer are shown in Table 1.

\begin{table}[h]
\begin{tabular}{cccccc}
\hline
Array configuration & No. of Inputs Nodes & $L1_N$ & $L2_N$ & $L3_N$ & No. of Trainable Parameters \\ \hline
Full Aperture       & 96            & 315    & 63     & 63     & 58,591                      \\
Pitch-Catch         & 256           & 354    & 55     & 55     & 113,911                     \\
Pulse-Echo          & 136           & 353    & 68     & 68     & 72,511                      \\ \hline
\end{tabular}
\title{Table 1: Network configurations showing the number of nodes for each layer including the three hidden layers (L1-L3) for the full aperture, pitch-catch and pulse-echo transducer array configurations.}
\label{table:dnn}
\end{table}

\begin{figure}[h]
    \centering
    \includegraphics[width=0.7\textwidth]{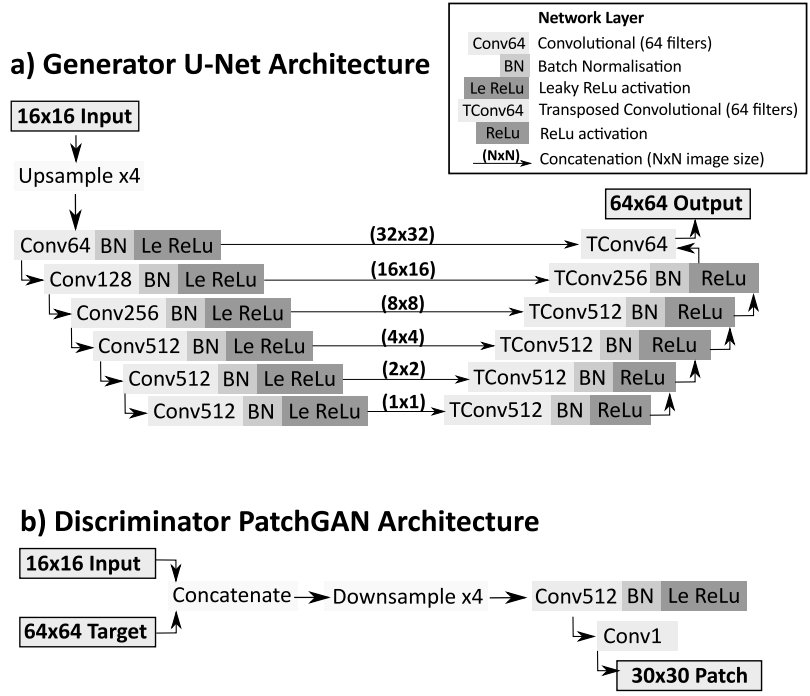}
    \caption{Network configurations for (a) the GAN generator U-net architecture and (b) the GAN discriminator PatchGAN architecture. All convolutional layers have a kernel size of 4}
    \label{fig:gan_arch}
\end{figure}

The GAN generator is a modified U-Net based on \cite{isola2017image} consisting of an encoder-decoder chain. Each block in the encoder is a convolution-batch normalisation-leaky rectified linear unit (ReLu) activation sequence. Each block in the decoder is a transposed convolution -batch normalisation-ReLu sequence with skip connections between mirrored layers in the encoder and decoder stacks \cite{ronneberger2015u} (as shown in Fig. \ref{fig:gan_arch}a). All convolutional layers use a kernel size of 4. The generator loss is the discriminator sigmoid cross entropy loss of the generates image with an array of ones combined with the mean absolute error between the generated and known target image.  

The GAN discriminator (Fig \ref{fig:gan_arch}b) follows a PatchGAN architecture \cite{isola2017image}, which divides the image into smaller 30x30 patches and the discriminator tries to classify each patch separately. This motivates the GAN to discriminate high frequency structure. The discriminator receives the target and generated images as well as the low resolution input.  The discriminator loss it the sigmoid cross entropy loss with the real image and an array of ones, combined with the sigmoid cross entropy loss with generated image and an array of zeros.

\subsection{Structural Similarity Index Measure (SSIM)}
We use the SSIM described by \cite{wang2004image} for image comparison. The SSIM is defined as a weighted combination of comparisons between image luminance $l(X,Y)$, contrast $c(X,Y)$ and structure $s(X,Y)$, where $X$ and $Y$ describe an image window in known and estimated images of size $N\times N$. The SSIM is therefore

\begin{equation}
    SSIM(X,Y)=[l(X,Y)]^{\alpha} \cdot [c(X,Y)]^{\beta} \cdot [s(X,Y)]^\gamma 
\end{equation}
where $\alpha$, $\beta$ and $\gamma$ are the weighting parameters. We use $\alpha=\beta=\gamma=1$. Luminance, contrast and structure are calculated as
\begin{equation}
    l(X,Y)=\frac{2\mu_X\mu_Y+C_1}{\mu_X^2+\mu_Y^2+C_1},
\end{equation}
\begin{equation}
    c(X,Y)=\frac{2\sigma_X\sigma_Y+C_2}{\sigma_X^2+\sigma_Y^2+C_2},
\end{equation}
\begin{equation}
    s(X,Y)=\frac{\sigma_{XY}+C_3}{\sigma_X\sigma_Y+C_3}
\end{equation}
where $\mu$ and $\sigma$ are the mean and variance of the windows X or Y and $\sigma_{XY}$ is the covariance of X and Y. This is computed over a sliding Gaussian window of $9\times9$.

\section{Acknowledgments}
This work was funded by the Engineering and Physical Sciences Research Council (UK): grant number EP/P005268/1.

\section{Data Availability}

The data and Python scripts required to reproduce these findings are available at:

\href{https://github.com/jonnyrsingh/DeepLearningAnisoTomo/blob/main/DeepLearningAnisoTomo.ipynb}{https://github.com/jonnyrsingh/DeepLearningAnisoTomo},
which can be executed within Google colabotory. This requires no additional software or downloads for the user. 

\singlespacing
\bibliographystyle{abbrvnat}
\bibliography{ndt_refs}

\end{document}